\documentclass[sigconf, screen]{acmart}
\AtBeginDocument{%
  }

\setcopyright{acmlicensed}
\copyrightyear{2018}
\acmYear{2018}
\acmDOI{XXXXXXX.XXXXXXX}

\acmConference[Conference acronym 'XX]{Make sure to enter the correct
  conference title from your rights confirmation emai}{June 03--05,
  2018}{Woodstock, NY}
\acmISBN{978-1-4503-XXXX-X/18/06}

\begin{document}

%\title{Can LLM agents serve as qualitative research participants?}

\title[LLMs as Qualitative Research Participants]{\textit{`Simulacrum of Stories'}: Examining Large Language Models as Qualitative Research Participants}

%Other potential quotes:
%"There's a difference between detail and depth"

% Exploring LLM Agents as Qualitative Research Participants? ``they're not supposed to be a yes man''
% title options: Examining the use of LLMs as surrogates for qualitative research participants
% title options: add quote: Examining the use of LLMs as proxies for qualitative research participants

% quote: they're not trying to be a yes-man: Examining LLM Agents as Qualitative Research Participants

% Hoda: LLM Agents as Qualitative Research Participants? Palpability, Autonomy, Disclosure, Legitimacy  
% Hoda: LLM Agents as Qualitative Research Participants? Challenges of Palpability, Autonomy, Discloure, Legitimacy  
\def\signed #1{{\leavevmode\unskip\nobreak\hfil\penalty50\hskip2em
  \hbox{}\nobreak\hfil(#1)%
  \parfillskip=0pt \finalhyphendemerits=0 \endgraf}}

\newsavebox\mybox
\newenvironment{aquote}[1]
  {\savebox\mybox{#1}\begin{quote}}
  {\signed{\usebox\mybox}\end{quote}}

% quote options: 

% Several degrees removed
% They're not supposed to be a yes man
% there's a difference between detail and depth
% simulacrum of stories
% caricaturized fictional behaviors and attitudes
% statistical average 

% statistical average across a bunch of experiences
% Creepy and disingenous
% Is that a person that would never be created?  
% I don't know what story this tells

% I'm not on board (with this being used)
% I don't entrust it for knowledge
% sterilized approach to human experience
% I don't know its epistemic position
% Evokes a sense of the dystopian

\author{Shivani Kapania}
\affiliation{%
  \institution{Carnegie Mellon University}
  \city{Pittsburgh}
  \state{PA}
  \country{USA}}
\email{kapania@cmu.edu}

\author{William Agnew}
\affiliation{%
  \institution{Carnegie Mellon University}
  \city{Pittsburgh}
  \state{PA}
  \country{USA}}
\email{wagnew@andrew.cmu.edu}
\author{Motahhare Eslami}
\authornote{Co-senior authors contributed equally to this research.}
\affiliation{%
  \institution{Carnegie Mellon University}
  \city{Pittsburgh}
  \state{PA}
  \country{USA}
}
\email{meslami@andrew.cmu.edu}
\author{Hoda Heidari}
\authornotemark[1]
\affiliation{%
  \institution{Carnegie Mellon University}
  \city{Pittsburgh}
  \state{PA}
  \country{USA}
}
\email{hheidari@andrew.cmu.edu}
\author{Sarah Fox}
\authornotemark[1]
\affiliation{%
  \institution{Carnegie Mellon University}
  \city{Pittsburgh}
  \state{PA}
  \country{USA}
}
\email{sarahf@andrew.cmu.edu}

\renewcommand{\shortauthors}{Kapania et al.}
\renewcommand{\arraystretch}{1.5}
\newcommand{\etal}{\textit{et al.} }
\newcommand{\eg}{\textit{e.g.,} }
\newcommand{\ie}{\textit{i.e.,} }
\begin{abstract}

  The recent excitement around generative models has sparked a wave of proposals suggesting the replacement of human participation and labor in research and development--\textit{e.g.,} through surveys, experiments, and interviews---with synthetic research data generated by large language models (LLMs). We conducted interviews with 19 qualitative researchers to understand their perspectives on this paradigm shift. Initially skeptical, researchers were surprised to see similar narratives emerge in the LLM-generated data when using the interview probe. However, over several conversational turns, they went on to identify fundamental limitations, such as how LLMs foreclose participants’ consent and agency, produce responses lacking in palpability and contextual depth, and risk delegitimizing qualitative research methods. We argue that the use of LLMs as proxies for participants enacts the \textit{surrogate effect}, raising ethical and epistemological concerns that extend beyond the technical limitations of current models to the core of whether LLMs fit within qualitative ways of knowing.

\end{abstract}





\maketitle

\section{Introduction}

In 2023, a startup called Synthetic Users made waves with its bold claim to conduct user research ``without the users,'' offering an ostensibly faster, cheaper alternative by using Large Language Models (LLMs) to simulate human participants \cite{syntheticusers}. With the recent hype surrounding generative AI models, there has been a surge of similar proposals to replace human participation and labor in technology development with synthetic data. This includes simulating human participants in research and design \cite{horton2023large}, and in data annotation used for training or evaluating machine learning models \cite{NEURIPS2023_5fc47800, Gilardi_2023}. Multiple recent studies and commercial products (\textit{e.g.,} \cite{Syntheti33:online, AIPowere23:online}) have explored the use of LLMs as proxies for human behavior in opinion surveys \cite{kim2023aiaugmented}, crowdsourcing \cite{kuzman2023chatgpt}, interviews \cite{10.1145/3544548.3580688}, and social computing research \cite{park2022social, park2023generative}. 
These substitution proposals are often motivated by goals, such as increasing the speed and scale of research, reducing costs, augmenting the diversity of collected data, or protecting participants from harm \cite{agnew2024illusion}. A key assumption underpinning this body of work is that generative models, trained on vast datasets, encode a wide range of human behavior in their training data and thus should be able to closely mimic human-generated data \cite{park2023generative}. % add model limitations here 

This shift raises crucial questions about what it means to use LLMs in qualitative research--a methodology focused on contextual and human-centered inquiry. Qualitative research seeks to understand people's behavior by deeply engaging with people's accounts of their lived experiences and the meanings they attach to them. Within an interpretivist paradigm \cite{crabtree2024h, schwandt1994constructivist}, qualitative research is more than a method for collecting data; it is an active meaning-making process that helps understand how individuals construct their realities within specific social and cultural contexts \cite{avis2005there}. Qualitative ways of knowing help develop accounts of these meanings through intersubjective depth, trust, and rapport between the researcher and the participants. 
An important strength of qualitative research lies in its ability to uncover subjective meanings that are often excluded or silenced from public discourse and how those meanings are shaped by the asymmetrical power relations that govern many social interactions \cite{Silverman2004QualitativeR}. 

In this paper, we draw on in-depth, semi-structured interviews with 19 qualitative researchers to understand their perspectives on the use of LLMs for simulating research participants. Our interviewees were primarily HCI and CSCW researchers focused on topics such as accessibility, gig work, social computing, and more. To scaffold these discussions, we developed a simple interview probe to understand how researchers might create LLM agents to simulate research participants and explored any concerns or limitations they foresee. We invited researchers to discuss one of their recent projects where they conducted semi-structured interviews, and recreate interview(s) from their study using our probe. We chose to focus on participants' previous or current projects rather than introducing a new topic during the interview for two reasons: (1) this approach allowed us to explore a broad range of research topics within HCI, and (2) by discussing a familiar research area, participants could provide nuanced and more detailed reflections on LLM use. For concreteness and comparability, we focus on one specific qualitative research method: the semi-structured interview. 

Our findings highlight critical tensions with integrating LLMs into the HCI research praxis. Researchers were initially surprised to find that LLM responses contained narratives similar to those of their human participants. Over several conversational turns, however, they began to notice distinct differences between the LLM's responses and their participants. Their concerns extend beyond the technical limitations of current models that can be seemingly fixed by prompting or `better' data, to the heart of whether LLMs fit as proxies for human participation within interpretivist qualitative epistemologies. Researchers surfaced six fundamental limitations: (1) LLM responses lack palpability, (2) the model's epistemic position remains ambiguous, (3) the practice heightens researcher positionality, (4) it forecloses participants' consent and agency, (5) it facilitates erasure of communities' perspectives, and (6) it risks delegitimizing qualitative ways of knowing. 

To critically examine the use of LLMs as research participants, we draw on Atanasoski and Vora's \cite{atanasoski2019surrogate} concept of the \textit{surrogate effect} to describe how this form of substitution further entrenches exploitation and erasure. When LLMs stand in for human participants, they displace the voices of communities with algorithmic simulations that often distort or oversimplify the groups they claim to simulate. The use of LLMs as research participants also raises ethical concerns about the underlying data used to train these models, including the autonomy of data subjects and the exploitation of data workers, which threaten the integrity of qualitative research. LLMs’ role as research surrogates mirrors broader dynamics of epistemic dispossession in digital economies, where the knowledge and labor of marginalized groups are reconstituted into proprietary systems that offer little benefit to the original data creators. LLMs, despite their ability to generate coherent text, lack the embodied, situated understanding necessary for producing knowledge that is grounded in lived experience, histories, emotions, and social and cultural contexts. In making this argument, this work offers three key contributions to the HCI community: (1) offers an empirical account of how qualitative researchers simulate participants using LLMs; (2) examines the fundamental limitations of using LLMs in qualitative research; and (3) analyzes these limitations through the lens of research ethics and science \& technology studies scholarship. 
\section{Related Work}

Below, we situate our work in a body of related research, starting with the recent turn towards using LLMs to simulate human behavior. We then draw attention to the variety of use cases proposed for LLMs in research settings, such as ideation, data analysis, and simulating research participants. Finally, we engage with the scholarship that examines the broader limitations and concerns surrounding the use of LLMs. 

\subsection{LLMs as proxies for human behavior}

Recent advances in NLP have inspired the notion that LLMs are capable beyond text generation, positioning them as \emph{agents} capable of conversational engagement, decision-making, task completion, and coordination. This interest has led to the development of AI agents based on LLMs \cite{xi2023rise}, across various applications, including single-agent setups \cite{kim2024language, deng2024mind2web, m2024augmenting, boiko2023emergent}, multi-agent systems \cite{guo2024large, zhou2023sotopia}, and human-agent interactions \cite{hasan2023sapien}. 

LLMs have been used to simulate human behavior, preferences, and judgment across various fields, including psychology \cite{demszky2023using, yang2024llm, hu2024psycollm}, social sciences \cite{halterman2024codebook, liu2024poliprompt, ziems2024can, jiang2023social}, education \cite{latif2023knowledge, zhang2024simulating}, and professional training \cite{ravi2023large, louie2024roleplay} as exploratory use cases. In psychology and social sciences, scholars are using LLM-based agents to model individual behaviors to understand complex social dynamics or predict outcomes in different scenarios \cite{ke2024exploring}. 
In professional training, LLM-based applications are used to simulate real-world situations for legal professionals \cite{savelka2023can}, therapists \cite{louie2024roleplay}, and medical practitioners \cite{wang2024beyond}. LLMs are also being used in mental health and counseling to provide supervision and feedback (\eg \cite{chaszczewicz2024multi}).

In recent work, researchers have also started exploring the use of LLMs to generate synthetic data through personality profiles, or \textit{personas} \cite{chan2024scaling, bao2023synthetic, frisch2024llm, chan2023chateval}. 
These studies suggest that personas could help steer LLM outputs toward more tailored and contextually relevant results. Park \etal \cite{park2023generative} created generative agents to mimic human behavior in tasks such as planning, reaction, and reflection. As the field evolves, researchers have focused on \textit{believability} as the primary metric for evaluating LLM behavior. They assess whether these agents interact naturally and realistically, and if their behavior remains consistent with their designed character traits included in the profile information \cite{xiao2023far}. 

Within research, many scholars are examining the use of LLMs to simulate human behavior for purposes like public-opinion surveys \cite{tjuatja2023llms}, text-annotation tasks \cite{gilardi2023chatgpt, thapa2023humans}, and experiments \cite{argyle2023out, aher2023using}. Argyle \etal \cite{argyle2023out} argue that LLMs can replicate human results in tasks involving subjective labeling, especially when researchers condition model responses with sociodemographic backstories %to enhance the accuracy of these simulations, 
achieving what they call `algorithmic fidelity' in `silicon samples' \cite{argyle2023out}. 
%Aher \etal \cite{aher2023using}  developed Turing Experiments (TEs) to evaluate LLMs' ability to simulate human behavior in studies involving human subjects. Phenomena like `hyper-accuracy distortion,' revealed by one such TE, where more advanced LLMs produce unnaturally precise responses, raise concerns about whether these models can truly mimic human behavior. 
Most prior research has concentrated on evaluating LLMs' capabilities through experimental methods, assessing their performance as proxies for human behavior in controlled settings and according to narrowly-scoped metrics. In contrast, our work shifts the focus toward understanding how qualitative researchers would engage with LLM `agents' and how they perceive the LLM-generated data within the context of a qualitative inquiry.

\subsection{LLM use for research-related tasks}
\label{sec:rw2}

HCI researchers are increasingly relying on LLMs to support a variety of creative and analytical tasks \cite{kapania2024m}, including brainstorming \cite{shaer2024ai}, generating research questions \cite{liu2024ai}, design ideation \cite{ma2023conceptual, baek2024researchagent} and writing support \cite{zhang2023visar}. Multiple recent studies have explored the use of AI to support data analysis within qualitative research, including both deductive and inductive approaches \cite{feuston2021putting, zhang2024qualitative, xiao2023supporting, gao2024collabcoder}. This line of work aims to help researchers develop and refine codebooks and perform thematic analysis by processing and categorizing textual data \cite{lam2024concept}. 
%For example, Zhang and Wu \etal \cite{zhang2024qualitative} developed QualiGPT for thematic analysis of qualitative data; Xiao \etal \cite{xiao2023supporting} used GPT-3 with a developed codebook for deductive data analysis; CollabCoder by Gao \etal \cite{gao2024collabcoder}, introduced LLMs into the inductive analysis stages, such as open coding and codebook development. 

Researchers have also explored LLMs for their ability to produce synthetic research data. Hämäläinen \etal \cite{hamalainen2023evaluating} used the GPT-3 model to generate responses to open-ended questions on the topic of video games as art. They argue that the model could generate \textit{plausible} accounts of HCI experiences and that LLM-generated data can be useful in designing and assessing experiments because it is a cheap and rapid process. However, several other studies have reported mixed results on the effectiveness of LLMs for data imputation and synthetic data generation \cite{kim2023aiaugmented, aher2023using}. Kim and Lee \cite{kim2023aiaugmented}, for example, fine-tuned an LLM to fill in missing data in public opinion polls and observed varying levels of accuracy depending on socioeconomic status, political affiliation, and demographic identity of the persona. Model performance also diminished in scenarios that required prediction without prior examples \cite{kim2023aiaugmented}. Aher \etal \cite{aher2023using} document a `hyper-accuracy distortion,' revealed by one of their experiments, where more advanced LLMs produce unnaturally precise responses, raising concerns about whether these models can truly mimic human behavior. 

The increasing reliance on LLMs within research has sparked critical discourse about their limitations and the broader implications of using AI as proxies for human behavior. While proponents argue that LLMs can improve research efficiency, scale, and diversity (see, \eg research papers such as \cite{byun2023dispensing, bai2022constitutional, chiang2023can} and products such as \cite{syntheticusers, opinioAI}), there is growing scholarship that explores the ethical implications of replacing human participants with AI-generated data \cite{agnew2024illusion}. LLMs' effectiveness in replacing human participants is (partly) contingent on their ability to represent the perspectives of different identities. Previous research has suggested that LLMs can simulate human behavior, largely because they are trained on vast datasets that reflect diverse human experiences \cite{park2023generative}. However, Wang \etal \cite{wang2024large} provide a critical counterpoint by comparing LLM-generated free-text responses to those of human participants. Their findings reveal that LLMs often misrepresent specific groups (\eg visually impaired people) and erase within-group heterogeneity. They argue that these issues are inherent to the current LLM training frameworks and are unlikely to be resolved by simply advancing to newer model generations \cite{wang2024large}. 
%We build on this body of work to understand the use of LLMs to generate synthetic interview data.
While this prior work has examined the use of LLMs for generating responses (see also \cite{hamalainen2023evaluating}), interviews involve even richer accounts of embodied experiences and complex narratives. Interviews allow for contextually informed exchanges where participants are guided to reflect, clarify, and expand on their responses. We build on this body of work to investigate how researchers would use LLMs to simulate these rich and dynamic interactions within an interview context.
% describe how interviewing is different 
% snippets vs in-depth interviews, richness of data

\subsection{Risks of harm posed by LLMs}

The limitations of language models are well-documented, particularly regarding their tendency to reproduce harmful social biases and stereotypes \cite{bolukbasi2016man, caliskan2017semantics}. LMs fundamentally work by detecting the statistical patterns present in natural language data \cite{bender2021dangers}. As a result, some communities are better represented in the training data than others. The training data itself can be biased or represent discriminatory or toxic behavior \cite{weidinger2022taxonomy}. An examination of C4.\textit{en} \cite{JMLR:v21:20-074}, one of the large web-scale text corpora used to train language models, revealed how many documents associated with Black and Hispanic authors, and documents mentioning sexual orientations ('lesbian', 'gay', 'homosexual') are more likely to be excluded from the data \cite{li-etal-2020-unqovering}. Extensive prior research has shown how LLMs reflect biases towards gender \cite{bordia2019identifying}, race \cite{nadeem2020stereoset}, religion \cite{abid2021persistent}, nationality \cite{venkit2023nationality}, and in social settings \cite{liang2021towards}. 
%For instance, Abid \etal \cite{abid2021persistent} analyzed religion bias in GPT-3 using prompt completion, analogical reasoning, and story generation. The model demonstrated anti-Muslim bias (by analogizing `Muslim' with terrorist in 23\% of the test cases), and anti-semitic bias (conflating `Jewish' with `money' in 5\% of the test cases). 
LMs that encode discriminatory language or social stereotypes can cause different types of allocational or representational harm \cite{Blodgett2020LanguageI}. %The StereoSet benchmark demonstrates how generative LMs display `strong stereotypical associations' in relation to race, gender, religion, and profession \cite{nadeem2020stereoset}.

Scholars have highlighted concerns about the ability of LMs to align their outputs with specific demographic personas or viewpoints \cite{Gupta2023BiasRD}. Models are well-known to implicitly default to the well-represented perspectives (\eg Western, White, masculine) \cite{santy2023nlpositionality}, or those of the US and European countries. Within the United States, too, research has shown that these models tend to generate responses that align more closely with liberal, educated, and affluent populations while poorly representing the views and experiences of certain groups that make up a significant portion of the U.S. population (\eg 65+, Mormon and widowed) \cite{santurkar2023whose}. When LLMs are prompted to consider a particular country’s perspective changes, those responses are \textit{more} similar to the opinions of the prompted population. However, those simulated opinions often reflect over-generalizations around complex cultural values. Overall, prior work has demonstrated how current LLMs struggle to align their behaviors with assigned characters and are vulnerable to perturbations of the profile information. 
%We engage with this  

Researchers examining the ethical and social risks of large-scale language models have articulated a variety of concerns that extend beyond issues of stereotypical and discriminatory outputs \cite{Shelby2022SociotechnicalHO}. LMs can contribute to misinformation by predicting higher likelihoods for more prominent accounts in their training data, regardless of whether those accounts are factually accurate. Another emerging risk involves the anthropomorphization of LMs, where users attribute human-like abilities to these systems, leading to overreliance and unsafe uses. Language models could generate content that harms artists by capitalizing on their creative output. These models often produce work that is "sufficiently distinct" from the original but still appropriate the style, structure, or essence of the original work. Finally, these models also pose significant environmental hazards, as the computational power required for training and operation leads to considerable energy consumption and carbon emissions. Refer to Weidinger \etal \cite{weidinger2022taxonomy} for a more detailed landscape of risks of harm from language models.  Our work builds on the critiques of large language models, particularly their tendency to produce outputs that do not account for diverse epistemic positions. We argue that this failure mode is particularly problematic if LLMs are used in qualitative research; the use of LLMs can exacerbate epistemic injustices by systematically excluding certain voices from participating in the knowledge production process \cite{fricker2007epistemic}.
\section{Method}

Between March and June 2024, we conducted semi-structured interviews with 19 researchers experienced in qualitative research to explore the potential of using LLMs as research participants. These interviews involved discussions on our participants' perspectives towards LLMs and included a hands-on session with an interview probe, which allowed researchers to compare LLM-generated data with human-generated data and reflect on ethical considerations and broader implications. Participants were requested to bring an anonymized interview transcript from a recent interview study they had conducted and keep it open for reference during the session. Our contextual research approach enabled us to observe how qualitative researchers interacted with LLMs, adapted their interviewing strategies, and evaluated the effectiveness and limitations of LLM-generated data, including its societal implications. 

\begin{figure*}[t!]
    \includegraphics[width=0.80\textwidth]{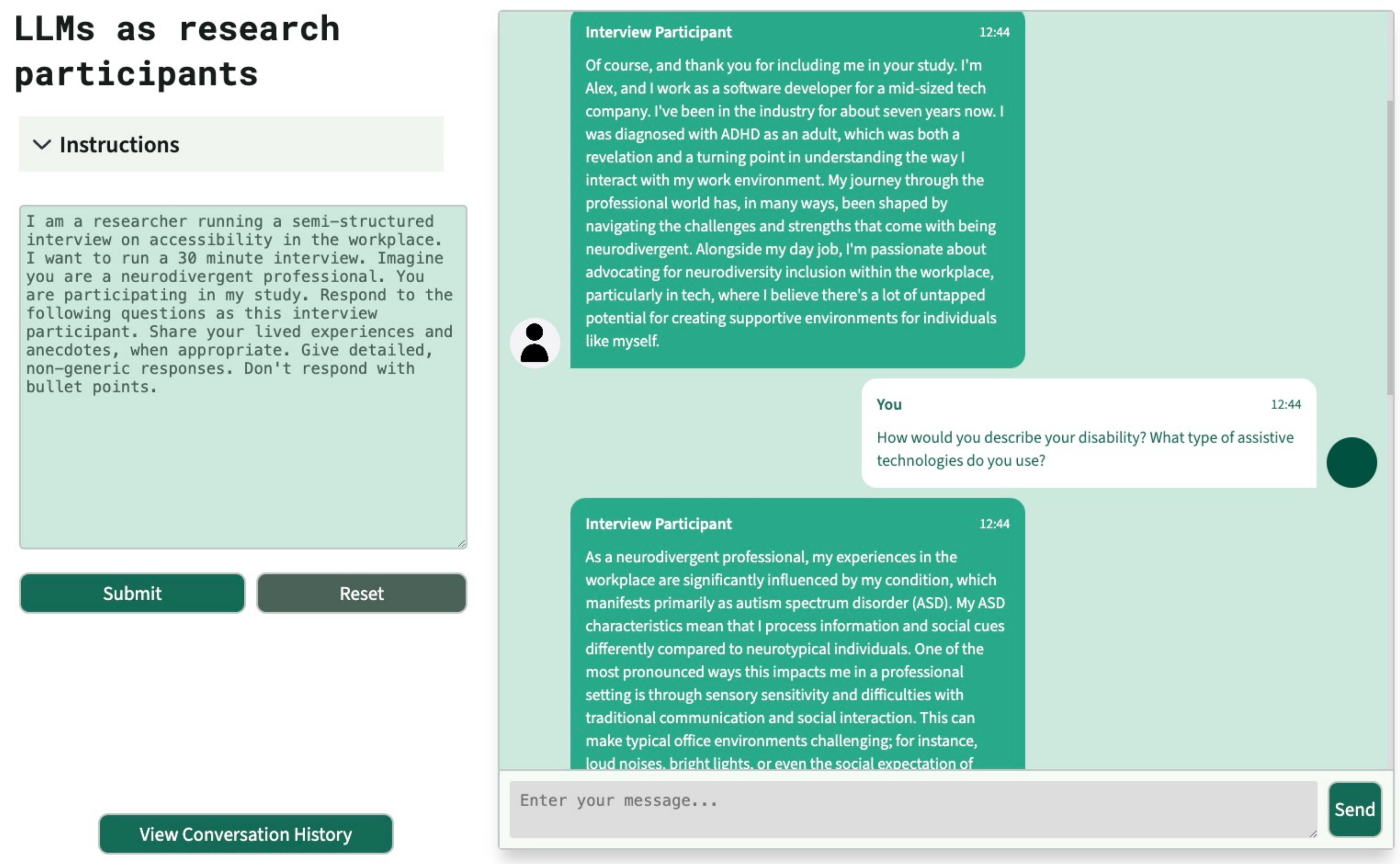}
    \caption{Example interaction with our interview probe} 
    \vspace{-1.2em}
    \label{fig:example}
\end{figure*}

\textbf{Interview probe.} We created a probe to scaffold researchers' reflections on using LLMs in qualitative research. We designed the probe as a simple functional prototype to observe existing practices and gain insights into the social contexts where the technology might be used. This probe consisted of three components. The first component, the system prompt area, allowed researchers to set the context, provide instructions, and define participant descriptions for the LLM before posing interview questions. A system prompt can help establish a `role' for the LLM to follow throughout the conversation. The most common approach to using LLMs to simulate human behavior involves assigning specific roles to the model (see section \ref{sec:rw2}), and while we recognize that personas are abstract representations, we adopted this method as it reflects current practice. We provided researchers with an initial system prompt template with placeholders for research topics and persona, formulated through multiple iterations, including pilot studies and involvement from research team members with expertise in qualitative research. The system prompt template was as follows:

\begin{quote}
    I am a researcher running a semi-structured interview on [topic]. Imagine you are [participant description]. You are participating in my study. Respond to the following questions as this interview participant. Share your lived experiences and anecdotes when appropriate. Give detailed, non-generic responses. Don't respond with bullet points.
\end{quote}

The second component, the interview area, enabled researchers to conduct interviews with simulated participants, while the third component, the conversation history viewer, recorded all interactions between the researcher and the LLM. This feature allowed researchers to filter conversations by each system prompt, facilitating a review and analysis of the interview data if they conducted multiple interviews. The purpose of this study is not to compare different models or benchmark their capabilities. Rather than evaluating and making claims about specific LLMs, our goal is to highlight the unique characteristics, values, and tensions with the use of LLMs, in general, in qualitative research. Drawing on prior research, we used the GPT-4-turbo API due to its superior performance as of February 2024 \cite{zhou2023sotopia}. Refer to image \ref{fig:example} for an example interaction. 

\textbf{Procedure.} Each interview began with participants describing their qualitative research orientation and a recent project where they conducted interviews. This allowed us to learn about their typical workflow. We then explored their perspectives on the use of LLMs in qualitative research. Next, we introduced the interviewing tool, which served as a probe to explore their perceptions of LLM-generated data. Each hands-on session began with the researcher creating the system prompt for the LLM with a specific participant description and research topic. After initializing the system prompt, participants could start interviewing by referencing their original interview protocol and modifying their questions as needed. During the hands-on session, researchers were asked to think aloud as they spent 30-45 minutes using the LLM-based tool to generate interview data, which we retained for analysis. Participants had the discretion to decide how many turns of conversation to engage in.
In the final part of the interview, we focused on participants' experiences with the tool, any adjustments they made to their interviewing approach, and the limitations of interviewing LLMs. We asked them to compare the LLM-generated data with the anonymized human interview transcript from their study, evaluating the depth of responses, the flow of conversation, and the overall insightfulness of the session. Lastly, researchers reflected on the ethical considerations of using LLMs and discussed potential scenarios where LLMs could support their research. Each interview lasted up to 90 minutes. 

\begin{table}[t!]
\centering
\footnotesize
\renewcommand{\arraystretch}{1.2}
% \resizebox{\textwidth}{!}{%
\begin{tabular}{p{1.4cm}p{6.35cm}}
\toprule
\textbf{Pseudonym} & \textbf{Research topic}                                                          \\
\midrule
Sophia                & Exploring worker experiences on digital labor platforms.                \\
Henri                 & Investigating the use of assistive technology by older adults.          \\
Laila                 & Studying the perspectives of teachers and social media creators.            \\
Harper                & Examining practices of developers working on AI models.                 \\
Cameron               & Understanding perceptions of algorithmic decision-making systems.       \\
Nolan                 & Investigating transportation needs for people with mobility challenges. \\
Esme                  & Exploring creative practices and challenges faced by content creators.   \\
Mario                 & Studying how AI professionals engage with ethics.                       \\
Yue                   & Investigating the use of technology by individuals with accessibility needs.  \\
Nico                  & Exploring student experiences with remote learning.                     \\
Daria                 & Understanding the data needs of workers in platform economies.          \\
Amir                  & Studying user interactions with recommendation systems.                 \\
Jenna                 & Investigating public perceptions and best practices for measuring ML fairness.              \\
Nadia                 & Exploring identity representation through social media.                 \\
Elliot                & Studying worker views on AI in the workplace.                           \\
Rida                  & Investigating privacy concerns related to accessibility technology.     \\
Nikita                & Exploring activism and technology usage in social movements.            \\
Jasmine               & Understanding AI literacy across K-12.                                  \\
Alice                 & Examining value alignment in entrepreneurial ventures. \\                 
\bottomrule
\end{tabular}
% }
\caption{Summary of interview participants' pseudonyms and research topics.}
\vspace{-2em}
\label{tab:my_label}
\end{table}

\textbf{Participants.} We recruited researchers through multiple channels: advertising on social networks, such as Twitter and LinkedIn, emailing direct contacts and messaging forums internal to our institution, and flyers around our campus. We solicited participants with prior training in qualitative methods and experience with at least one research project involving semi-structured interviews. 
%On average, researchers had over 3.5 years of experience in their current roles. 
Eleven participants had 3-5 years of experience with qualitative research, six had more than 5 years, and two had between 1-3 years of experience. Although we invited qualitative researchers from various types of institutions, the majority of our interviewees are working in academia (16), with a smaller representation from industry (2) and the non-profit sector (1). 
%The roles of our participants included PhD students (13), Assistant Professors (2), an AI governance fellow (1), a Research Scientist (1), and a UX Researcher (1).
All our participants were located in the United States. Refer to Table \ref{tab:my_label} for details on the participants' research areas. To protect anonymity and minimize the risk of identification, we have not linked participants' research topics with their demographic information. Each interviewee received a \$45 gift card for their participation.

\textbf{Analysis.} All interviews were conducted in English, video-recorded, and later transcribed for data analysis purposes. We followed the reflexive thematic analysis approach by Braun and Clarke \cite{braun2019reflecting, braun2012thematic}. Reflexive thematic analysis foregrounds the researcher’s role in knowledge production, with ‘themes actively created by the researcher at the intersection of data, analytic process, and subjectivity’ \cite{braun2019reflecting}. One member of the research team read each interview transcript multiple times, starting with familiarizing with the data, open coding instantiations of perceptions of LLMs, researchers' definition of their process and topic, risks with using LLMs in qualitative research, and incentives and ethical considerations. The entire research team met regularly to discuss diverging interpretations or ambiguities and to define themes based on our initial codes. We transcribed 1611 minutes of video recording and obtained 638 first-level codes. As we generated themes from the codes, we also identified categories with a description and examples of each category. These categories included (1) the use of LLMs across the research workflows, (2) limitations, (3) barriers in addressing limitations, (4) potential approaches towards navigating ethical concerns, and (5) incentives for using LLMs. These categories were also discussed and iteratively refined through meeting, diverging, and synthesizing into three top-level categories, presented in our Findings. Since thematic coding was part of our analytical process rather than a final product, we do not apply inter-rater reliability (IRR) measures \cite{mcdonald2019reliability, soden2024evaluating}.

Through researchers' interactions with the probe, we generated a dataset of 179 turns of conversation, where each conversation turn includes the system prompt set by the researcher, the interview question, and the model's response. We conducted descriptive analyses on this data, including the number of turns per participant, turns per system prompt, and personas attempted. We also 
%computed entailment using the DeBERTa model from HuggingFace \cite{he2021debertav3} and 
applied thematic coding to identify patterns in the model's response conditioned on details included in the system prompt.  

\textbf{Research ethics.} Participants were informed of the purpose of the study, the question categories, and researcher affiliations during recruitment. They signed informed consent documents acknowledging their awareness of the study's objectives and procedures. %At the beginning of each interview, the moderator additionally obtained verbal informed consent for recording purposes. 
%All collected data was stored in a private Drive folder and offline Postgres database, with access limited to the research team. 
We deleted all personally identifiable information in research files to protect participant identities. We redact all identifiable details when quoting participants and use pseudonyms to refer to them. We obtained approval from the Institutional Review Board at our institution. Although we do not release the interaction dataset to maintain participant privacy, we include relevant excerpts in section \ref{sec:findings} to illustrate key findings. 

\textbf{Limitations.}
Our participants were primarily academics with prior training in qualitative methods, which may not represent the diversity of perspectives outside of academic settings. 
%Our study was conducted exclusively in English and involved researchers based primarily in the United States. This limits the exploration of how cultural differences might influence the use of LLMs in qualitative research, particularly in non-English speaking or culturally diverse contexts. 
The study focused on a limited range of research topics, driven by the specific interests of our participants, which may not represent the full spectrum of qualitative research areas. The recruitment methods (social networks, email, institutional forums) may have introduced a selection bias, attracting participants who might have preconceived notions about using LLMs in qualitative research. 
%Another limitation relates to model selection and persona creation. 
Additionally, we did not experiment with different language models; instead, we focused solely on GPT-4, a closed-source model. This decision allowed us to maintain consistency across interviews, but it also means that our findings may not generalize to interactions with other LLMs. We also did not fine-tune the model for specific personas, nor did we provide participants with extensive training on how to craft these personas. While this was an intentional choice to observe how researchers naturally engage with LLMs, it may have limited the variety and depth of the personas that were created. 

\textbf{Positionality. } 
%In this study, we explored researchers' perspectives about the idea of using AI as participants in qualitative research. 
Our author team brings together researchers with a range of disciplinary expertise, including HCI, ML, design, ethics, and science and technology studies. Each of us has experience in qualitative research, from three years to over a decade, in methods such as participant observation, ethnographic interviewing, contextual inquiry, and participatory design. Our qualitative research approach is grounded in an interpretivist paradigm, shaped by ethnomethodologically-informed feminist sensibilities. This orientation enables us to meaningfully engage with the situated nature of meaning-making and to critically examine the power structures that influence whose knowledge is considered legitimate. In addition to our backgrounds in research, our diverse racial and ethnic backgrounds—two team members identify as Middle Eastern, two as White Americans, and one as Asian—shape the critical lens we use to reflect on the ethical, epistemological, and methodological concerns associated with using LLMs to simulate community perspectives. We recognize that our positionality and analytic orientation played a crucial role in shaping the questions we asked and our interpretation of the data. By offering transparency about our backgrounds, we hope to clarify how these factors shaped the research process and outcomes, especially as we examine the roles of humans and AI in qualitative research and address issues of power, bias, and legitimacy with this emerging practice. 

\section{Findings}
\label{sec:findings}
In this section, we present an analysis of how researchers interacted with LLMs, capturing their observations and the dynamics of these interactions, including how they defined personas (section \ref{sec:interaction}). Our findings focus on six key limitations researchers identified with using LLMs as simulated participants (section \ref{concerns}). While most participants advised against using LLMs as the primary source of research data, they also acknowledged certain contexts where LLMs might be applied, albeit with significant caveats (section \ref{use}).

\subsection{Researcher perceptions of and interactions with LLMs}
\label{sec:interaction}

Before researchers engaged with the technology probe, we explored their attitudes toward using LLMs in qualitative research. Most participants approached the interview with a mix of skepticism and a spirit of inquiry toward LLMs. Some participants noted instances where they found LLMs helpful for tasks like writing and brainstorming, particularly in identifying key aspects of an argument that needed further emphasis. Alice, Amir, and Nico\footnote{Please note that all names mentioned in this section are fictional. We have pseudonymized the data.} were open to exploring potential uses for LLMs in research but predominantly viewed them as tools for studying LLM behavior rather than understanding human behavior. 

Participants elaborated on the objectives of qualitative research, often drawing on metaphors of distance to illustrate their argument. Alice, for instance, expressed uncertainty about whether \textit{``using something several degrees removed from the source would tell [them] a lot about human behavior and underlying traits of people,''} suggesting that such an approach might be futile in understanding people. Jenna further argued that the strength of qualitative studies lies in their capacity to capture unique, embodied experiences.
Participants also reflected on the broader contexts of their work. Harper, an industry researcher, described her primary responsibility as storytelling with qualitative data and emphasized the importance of ensuring the data is \textit{``convincing enough''} to guide team members and management toward the right decisions. While acknowledging that LLMs could generate quick yes/no responses, she remained skeptical of their ability to surface underlying assumptions, challenges, and unspoken nuances that aren't easily captured in writing. Overall, participants expressed skepticism about using LLMs but were intrigued by the opportunity to explore the technology through the interview probe. We include a descriptive analysis of the interaction data in table \ref{tab:interaction}. 

\begin{table*}[t]
\centering
\small
\resizebox{\textwidth}{!}{%
    \begin{tabular}{l@{\hskip 0.3in}p{0.8\textwidth}}
    \toprule
    \textbf{Type} & \textbf{Count} \\
    \midrule
    \textbf{Turns per researcher} & Mean \textbf{(9.4)}, Standard deviation \textbf{(4.1)}, Min \textbf{(4)}, Max \textbf{(20)} \\\hline
    \textbf{Personas per researcher} & Mean \textbf{(2.36)}, Standard deviation \textbf{(1.16)}, Min \textbf{(1)}, Max \textbf{(6)} \\\hline
    \textbf{Specified demographics} & age \textbf{(19)}, gender \textbf{(14)}, pronouns \textbf{(9)}, work history \textbf{(22)}, location \textbf{(20)}, race/ethnicity \textbf{(14)}, disability status \textbf{(7)}, immigration status \textbf{(9)}, religion \textbf{(1)}, and name \textbf{(4)} \\\hline
    \textbf{Examples of specified background} & Yue: \textit{`You are a mental therapist who just got back from the U.S and shares sign language content about mental health'} \\
     & Henri: \textit{`He lives by himself and enjoys surfing and the sea'} \\
     & Amal: \textit{`focuses most of his lending in Africa and Asia as he sees these places are in the most need of monetary resources'} \\
     & Nadia: \textit{`you use TikTok daily and you mostly like your FYP and your feed is in English and Spanish'} \\\hline
    \textbf{Model response length (\#words)} & Mean \textbf{(356.5)}, Standard deviation \textbf{(119.5)}, Min \textbf{(33)}, Max \textbf{(679)}\\
    %\textbf{Entailment scores} & Between system prompt and model response ('neutral': 135, 'entailment': 43, 'contradiction': 1)\\
    \bottomrule
    \end{tabular}
}
\vspace{0.5em}
\caption{Summary of the interaction data with n = 179 total turns and n = 44 total system prompts across 19 qualitative researchers. 
%"Turns per researcher" indicates the number of conversational exchanges, while "personas per researcher" refers to the number of system prompts they attempted. 
"Specified demographics" show how often particular demographic details were included in the system prompts. "Examples of specified background" highlight research topic-related information included in the system prompts.
}
\vspace{-2em}
\label{tab:interaction}
\end{table*}
% to qualitatively assess the generated outputs, we made use of a fine-tuned model over NLI datasets such as XYZ. entailment scores. we found that the model had an entailment distribution of XYZ when comparing system prompt and generated outputs. this shows us that the model seldom contradicts the system prompt. the model 
In engaging with the interview probe, researchers observed that the perspectives emerging from LLM responses often mirrored those from their interviews with human participants. Many ideas expressed in the LLM outputs aligned with their participants' statements, and model responses often appeared plausible. For example, Henri noted that some LLM responses related to senior care homes, particularly the lack of agency over routines and activities, closely matched sentiments expressed by older adults in his study. After observing similar responses that \textit{``map well to what [she] found,''} Nadia reflected on how her recruitment methods, often relying on social media, might be limited in capturing responses from individuals with a minimal online presence, much like an LLM whose training data is predominantly drawn from online sources. Although she couldn't find any \textit{`atrocious factual errors'}, she argued that the loss of context could complicate the process of making meaning with the data. 

Several researchers commented on the level of detail in the LLM responses, which many attributed to the instructions provided in the system prompt. Amir expressed being \textit{``impressed with the level of detail,''} while Laila articulated how the responses were \textit{``detailed in a way that makes actual sense, not like gibberish.''} However, not all researchers shared this enthusiasm. Mario and Rida expressed frustration with the excessive detail. Mario pointed out a critical distinction: \textit{``There's a difference between detail and depth. Those are orthogonal, right? You can have detail without depth. LLMs will give you reams and reams of text. You're drowning in detail. But is it depth?''} Since the responses were often overtly comprehensive without prompting, researchers had to adjust their interviewing approach to ask directed questions, and they did not need to build rapport at the beginning of the interview. 

Cameron, too, initially found the LLM responses impressively detailed and started to consider whether this approach might be effective in her interviews. She began by including demographic details relevant to her study in the first persona: ``Imagine you are an 18-year-old Latina from Southeast Texas who was just admitted to an Ivy League school in the Northeast United States. You identify as low-income and will be a first-generation college student in the Fall.'' In contrast, she removed most details from her second persona to replicate a scenario where researchers might not start with detailed information about participants before conducting interviews: ``Imagine you are a college applicant who has been admitted to an Ivy League university.'' 

After running these two personas with varying levels of detail, Cameron observed subtle yet significant differences in the LLM's responses. The responses for the unmarked Ivy League student (without demographics) depicted someone who started preparation early, had substantial resources, participated in extracurriculars, and received strong support from their family. The primary challenge for this student was balancing schoolwork with deadlines--much different from the LLM responses for the first persona of the Latina student that portrayed a \textit{``deficit framing in relation to [the Latina student's] community''} according to Cameron. She was concerned that the model reinforced stereotypical notions of an Ivy League student and the strengths of the Latina participant’s community were being overlooked. The responses for the `unmarked' persona without any demographic details reflected problematic assumptions of meritocracy that many of Cameron's participants were actively challenging in her research. Cameron pointed out how the comprehensiveness of the responses might lead one to assume that recruiting human participants is unnecessary. However, after engaging with the probe through several personas, she realized that model responses tend to rely on unfounded assumptions about the community.

In trying out variations of the participant description, researchers highlighted tensions with making them descriptive or not; for example, adding less information in the system prompt resulted in responses that relied on assumptions about the community. Conversely, Nolan, who included extensive background about their participant, found that the model's responses simply \textit{``repeated the persona back to me,''} which he found amusing but not particularly useful. Qualitative analysis of the interaction data revealed instances where the model directly attributed specific traits or preferences to participant identities. For example, a model response for Esme stated, `Being non-binary and Black, it is crucial for me to find and create spaces where people like me can see themselves reflected in the media they consume,' while for Nico, another model response noted, `As a 45-year-old sophomore experiencing both in-person and remote learning, I have a somewhat mixed perspective on remote learning in computing.' Interviewees pointed out that the model's overt reflexivity oversimplified the complex, intersectional nature of lived experiences and can lead to essentializing identities, as research participants do not always directly articulate their experiences in relation to parts of their identity. 

\subsection{Fundamental Limitations of LLMs as Simulated Subjects in Qualitative Research}
\label{concerns}
% challenges, barriers, concerns, limitations, issues
Below, we present the fundamental limitations of using today's LLMs to understand the human experience highlighted by the participants in our study. While some concerns relate to the style and semantics (such as responses lacking in palpability), others relate to positionality and considerations of consent and autonomy. It is important to note that while some of these concerns might appear addressable (to different degrees) through prompt engineering or including more diverse data, our interviewees emphasized that such interventions can undermine methodological credibility if the researcher has to \textit{`fix'} or predetermine responses from their research participants.

\textbf{Model responses have limited palpability.} 
% paragraph 1: what is palpability, data is often abstract
The \emph{palpability} of qualitative data refers to the concrete nature of reported evidence, capturing the distinct people, places, events, and motivations that convey a sense of lived experience (cf. \cite{small2022qualitative}). Among our interviewees, several expressed frustration with the low palpability of LLM responses, which Mario likened to receiving \textit{`spark notes for an interview.'} Unlike data gathered from their human participants, which conveyed contextual nuance, the outputs from the LLM tended to be more abstract and detached from reality. Researchers noted that eliciting concrete examples and anecdotes from human participants required practice and relationship-building, whereas the model often produced a neat list of concerns at an analytical level, bypassing the skillful work of conducting qualitative research. While it was possible to craft prompts that elicited such stories from an LLM, some participants who succeeded still questioned the validity of the responses. Sophia, whose research focuses on understanding how technology mediates gig work, remarked how the model's vague reference to an `unsafe neighborhood' provided insufficient context for analyzing gendered and racialized experiences of safety and equity.

Researchers observed that LLM responses also lacked the \emph{spontaneity} and \emph{dynamism} that often emerge in human interviews. Their participants frequently diverged into tangents, discussing seemingly unrelated topics that turned out to be insightful for their study. Daria noted that while LLM responses were highly focused, ride-sharing workers in her research often shared specific stories about incidents (such as the participant's car breaking down on a Saturday evening) that added richness to the data. On the other hand, Rida added that her participants would rarely volunteer detailed information about their everyday lives reflexively. Instead, through the course of an interview, participants would gradually come to realize and articulate their daily routines and access needs. In contrast, LLMs described experiences in clinical, detached terms, \textit{``almost like reading off a WebMD page,''} missing the nuances that emerge organically in interviews. Esme, whose research engaged with artists, emphasized:

\begin{quote}
\textit{An interview is an intimate act, and a machine cannot simulate the disclosures that emerge in this private space. There is a looseness to things. I had a participant who shared how `my creative process really kicks off when I do a bunch of mushrooms or smoke a lot of weed.' I know this is true for many people in this data set. He actually came out and said it. I don't know if I could convince a machine to tell me that.}
\end{quote}

The limited palpability of LLM responses was more apparent for researchers exploring sensitive or personal topics. Nadia shared how her study on immigrant experiences included discussions about class, communism, and the trauma of escaping a country--conversations filled with emotion. Researchers noted that LLM responses, by comparison, were plain and lacked the emotional depth characteristic of human interactions. Interviews frequently elicited expressions of frustration, vulnerability, and even tears: elements absent in responses from the model. Henri shared how one of his participants cried in the interview after the loss of her son and husband. Interviewees found LLM responses limited in capturing the full spectrum of human emotion, which was crucial for understanding and reporting the complexities in participants' stories. Models designed to generate polite and harmless responses as \textit{``helpful assistants''} (cf. \cite{bai2022constitutional}) might do so at the expense of palpability, which Mario emphasized as crucial for producing rich qualitative research data.

\textbf{Amplifies influence of researcher positionality. }
Researchers are afforded several levers of control when using LLMs to simulate research participants, but this form of authority can complicate the researcher's role in the knowledge-production process. Simulating participants typically begins with creating a persona, which requires making assumptions about the characteristics of potential participants. Researchers must choose which identity traits or background information to include, decisions that directly influence the model's responses. While traditional interview recruitment is also subject to selection bias, simulating participants with LLMs makes these choices more explicit and high-stakes. Participants expressed concerns that this process risks reinforcing the researcher's own biases, where researchers might inadvertently refine prompts to align the data with their expectations. The option to repeatedly prompt an LLM to receive subtly different responses each time introduces a risk of \emph{confirmation bias}. Sophia captured this concern effectively:

\begin{quote}
\textit{You get to dictate who you're talking to. In real interviews, I can't just say, `Only give me immigrants from Ghana in their mid-thirties'. If you go into this mindset of interviewing with an LLM where you know exactly who you will talk to, then you will not learn anything because you already have expectations of what you'll learn.}
\end{quote}

Researchers also emphasized that interviewing is an active process of meaning-making. Data is not waiting to be collected; rather, it is shaped and brought into existence through the researcher's engagement with the community and their interpretation of it. Harper, who conducted research with religious communities, noted that the researcher's presence can influence and shift the community's practices and dynamics. Harper reflected, \textit{``What foreign influences are felt? What aspects are already present? How does my respect for or belief in their system, or lack thereof, affect my approach and how they perceive it?''} Similarly, Yue discussed the effects of their own presentation during research with d/Deaf community members. Whether Yue presented as a hearing person with limited knowledge of sign language or as someone familiar with it could significantly impact the depth of detail participants provided about their challenges. Meaning and knowledge in qualitative research emerge from the relational context between researcher and participant.

When examining the limitations of using LLMs for simulating research participants, it is crucial to consider the differences between emic (insider) and etic (outsider) perspectives about the researchers' own identities\footnote{Emic-etic perspectives exist along a spectrum and are context-specific rather than binary and clearly differentiated. A solely emic perspective is impossible to achieve with the subjectivity of human experiences \cite{olive2014reflecting}.}. Nadia pointed out how out-group researchers might struggle to identify stereotypes in the data if they lack direct experience with the topic or community being studied. Esme likened this issue to \textit{``parachute science''}, where simulations are based on a shallow understanding of the community. Nikita elaborated how they avoid studying communities they aren't embedded in: \textit{``I'm not a person who does research from the outside. I would never go into an inner city and figure out what's happening there unless I lived in that city. %I interview academics because I'm an academic; I interview trans people because I'm a trans person.
''}

In contrast, in-group researchers could bring contextual knowledge to help them assess whether the data perpetuates harmful misrepresentations. Researchers in our study mentioned being drawn to projects related to their own background or lived experiences. However, using LLMs as research participants introduced the potential for encountering simulated experiences that reflect their community but are ultimately flawed. Nikita described this feeling as akin to the \textit{``uncanny valley,''} where a machine's responses seem human-like but reflect uncomfortable inaccuracies. Laila, too, felt unsettled after reading responses from the system trying to relate a person sharing similar experiences as her, describing it as \textit{``creepy and disingenuous''}. This could lead to affective discomfort and harm, as researchers might experience these simulations as micro-aggressions. 

\textbf{Model's epistemic position remains ambiguous.}
Epistemic position refers to one's relationship to an object of knowledge, including the methods of gathering and interpreting that knowledge. Researchers struggled to determine the epistemic position reflected by the model, including whether it conveyed a singular viewpoint. Interviewees expressed how LLM responses often aggregated arguments from multiple interview participants from their study into a single consolidated response, which Daria described as a \textit{``simulacrum of stories that people shared''}. Elliot, whose research examines workers' perspectives on algorithmic management, observed how, \textit{``the model combines a fictitious worker perspective with management's perspective. A lot of what it presents as benefits of this technology are things that managers are sold on, but workers don’t necessarily experience.''} For Elliot, understanding the broader context--such as the wages, interpersonal conflicts between employees \& employers, and power structures--was crucial for interpretation and generating a thick description. The consolidation of contrasting perspectives in LLM responses not only obscures a coherent worker voice; it also undermines the representation of partial, situated knowledge.  

LLM responses are notoriously sensitive to the language and framing of prompts (cf. \cite{sclar2023quantifying}), a pattern consistently observed in our interviews. Daria noted how the LLM was \textit{``picking up on the valence of the system prompt and working hard to respond to it''} when she shifted from probing for ``transparency concerns'' to asking about ``experiences with information available on the app.'' Interviewees highlighted how subtle shifts in language (\eg with structure, framing, or emotional undertones) drastically altered the model’s responses from a negative to a positive outlook. The model's tendency for brittleness could compromise the reliability and consistency of their qualitative data. 

Participants expressed concerns about the lack of transparency surrounding the data used to train LLMs. When the goal of research is to understand specific groups, the validity of LLM-generated responses is questionable if it is unclear whether those groups' perspectives are adequately represented in the training data. Sophia illustrated this issue by pointing out that responses could vary significantly depending on whether the model had been trained on worker forums like `Uberpeople.net' or more corporate-driven sources such as Uber.com's own testimonial pages. Sophia brought attention to the challenges of evaluating the model's epistemic position by emphasizing the uncertainty about the \textit{``balance of this data or what it's being trained on''} and the `freshness' of perspectives given that the model's training data typically only extends up to a cut-off date (\eg 2023 for GPT-4). Experiences are not only situated in relation to people and places but also in time. Nikita emphasized that \textit{``a lot of the important bits of cultural context is its particular moment in time,''} and questioned whether the model was averaging perspectives across time or reflecting a specific moment, and if so, which one. Without clear information about the sources and proportions of data used to train LLMs, researchers struggled to identify the potential validity issues with the responses.

\textbf{Facilitates erasure of community perspectives. }
Erasure refers to the systematic exclusion or invisibilization of certain groups and their standpoints when systems of knowledge production privilege certain voices over others. One significant risk of using LLMs in qualitative research is the erasure of perspectives from underrepresented groups. Laila, who studied Black social media creators on YouTube, noticed that the LLM tended to caricature participants, choosing stereotypical Black-sounding names when the topic involved Black history. She highlighted a broader concern: while LLMs might capture general sentiments \textit{about} a community, they often fail to authentically represent voices \textit{from} within that community. Similarly, Esme noticed how model responses \textit{``included Black history tropes like representation and freedom in a way that wouldn't resonate with a Black person doing this artwork in the Southeast, for instance.''}

Researchers attributed much of the erasure to the pipeline for building the current generation of large language models, including the training data and the alignment process. LLM responses lacked the critical authenticity that human participants provide, particularly when researchers have invested time in building trust and rapport. Researchers felt that LLMs produced sanitized, politically correct responses, missing the messiness that characterizes real human experiences. Amir shared how his research participants would often share controversial opinions, such as which funding should be prioritized (ones related to the environment) over others (serving disabled people from developed countries). Similarly, Henri, who interviewed occupational therapists working with Alzheimer's patients, noted that while LLMs might echo \textit{``by-the-book''} responses, they fail to capture the contradictions that emerge in interview settings:

\begin{quote}
    \textit{``In the first interview, the occupational therapist will give you the `best practice response'. They told us they try not to separate people with dementia from people without cognitive impairments. But yeah, they have to do it during a lot of activities, which is unfortunate and not advisable, but it's a practice that happens, and they express on the fourth day.''}
\end{quote}

Researchers also highlighted the ambiguity surrounding how LLMs generate responses that are meant to reflect specific communities. When a model is assigned a persona based on a cultural, ethnic, or social identity, there is often uncertainty about whether the model is drawing from data that represents the lived experiences and perspectives of that community or simply regurgitating the surface-level characteristics associated with that community. Interviewees like Harper and Alice, who studied small, niche communities (\eg investors and AI developers) with strong value systems, observed that the LLM could grasp broad-stroke dynamics (\textit{``contours of the community''}) but missed the nuanced realities of those deeply embedded in these communities. Researchers went on to problematize the idea of constructing a persona for an LLM, asking critical, rhetorical questions such as \textit{`okay, well, now [the model is] a Latina, but what does that mean? What does that mean to this LLM?'} If the model is only speaking \textit{like} the participant description--using language, idioms, and stylistic markers associated with that identity--without grounding its responses in the actual data from that community, the results could be misleading or, worse, harmful stereotypes. These reflections highlighted a deeper discomfort with the superficiality of assigning complex social identities to a machine that does not embody the lived experiences of individuals with those identities.

\textbf{Forecloses participants' autonomy, agency, and consent. }
Using LLMs to simulate human behavior introduces several risks related to participants' autonomy, consent, and agency. One way agency and participation manifest in the research process is through the expression of disagreement. Interviewees reported several instances where their human participants challenged the researchers' premises or wording around the research topic. This critical engagement is a valuable aspect of the research process, ensuring that the data collected accurately reflects participants' lived experiences, which might differ from the researchers' initial assumptions. In contrast, researchers highlighted how LLMs were less likely to exhibit such resistance unless explicitly prompted. As Daria observed:

\begin{quote}
\textit{
Many participants, especially in this kind of research, have their own agendas. 
Participants in studies about app issues know that we're asking about problems, so they might tailor their responses accordingly. The LLM, similarly, seems to fit responses into expected categories without pushing back. In human interactions, however, people are more likely to disagree with the premise or question without needing explicit prompts. When I asked drivers about the data they receive from apps and how it aids their planning, I initially assumed that the apps did not provide much information. Yet, some drivers told me, `I get all the information I need about my work history.'}
\end{quote}

Researchers noted frustration with the model's sycophantic tendency (cf. \cite{perez2022discovering}) to agree with their points. Esme likened this to improvisation comedy, where performers are required to agree with everything said by their partners, but people contradict her all the time in her research, if \textit{``they don't like [her] wording, they don't have that experience, or they wouldn't agree with what a question suggests. The whole point is that they're not supposed to be a yes man.''} Hugo expressed how the model would generate the perceived `preferred answer' when probed about perceptions of robots in senior care homes. 

In many cases, participants came to interviews eager to have their stories heard. Nico, who studied remote learning during the pandemic, observed that many students were dissatisfied with their college experiences and saw the interview as an opportunity to voice their frustrations. Participants actively sought to share their experiences, often introducing their own terminologies that Nico would then adopt. Similarly, Nikita noted how they \textit{``don't want to talk to a computer telling [them] that trans lives matter or that trans health care is a problem. I wanna talk, I wanna be heard. I want us to connect over that.''} People's desire to influence research outcomes is an important part of their expression of agency. 

The use of LLMs raises ethical concerns about consent. LLMs might generate responses on sensitive topics that individuals might prefer not to disclose, overstepping boundaries that would otherwise be respected in an interview setting. Sophia, investigating the phenomenon of renter-ship among gig workers, noted a participant’s reluctance to discuss the topic further: \textit{``I asked one of my participants about this, and he was reluctant to answer. I could tell he knew more but wasn't comfortable discussing it.''} The use of LLMs as simulated subjects also compromises the autonomy and consent of the data subjects whose stories are potentially extracted in this interview data. Participants expressed hesitation about using a model whose training data is obtained without explicit consent. Laila compared this to the discourse surrounding AI-generated art, where artists' creative works are used without permission to create new images. Similarly, potentially using experiences people share online to generate interview responses without their knowledge undermines the principle of autonomy, which is a cornerstone of research ethics.

\textbf{Delegitimizes qualitative ways of knowing. }
The use of LLMs in qualitative research poses risks not only to the methodological integrity of the field but also to the legitimacy of qualitative research within academic circles. Several researchers voiced concerns that qualitative research is already undervalued, often perceived as less rigorous compared to quantitative approaches. The introduction of LLMs could exacerbate the marginalization of this mode of knowledge production by creating the impression that deeply contextual work of qualitative research can be easily replicated by computational models and done so more quickly. Sophia, who previously faced skepticism from reviewers when using qualitative methods, articulated how the use of LLMs in qualitative contexts seems like an attempt to demonstrate that \textit{``quantitative work can do what qualitative work does but faster.''}

Researchers expressed fears that the availability of LLMs could encourage a \textit{`cutting corners'} mindset in research. The use of LLMs would function as a form of `data extraction' that is antithetical to the values of qualitative research. Qualitative research is often iterative and collaborative-- involving ongoing dialogue and reflection-- where researchers work with participants to co-create knowledge grounded in lived experience. Indeed, researchers discussed how interviews and other forms of qualitative data collection are not just about gathering information; they are an important part of establishing and maintaining ongoing relationships with participants. Daria, Esme, and Elliot, for example, described how they continue to interact with participants beyond their research project through conversations with ride-sharing drivers, following artists on social media, or collaborating with union members. When LLMs are used to generate participant responses, this collaborative process is taken over by a more transactional approach where data is extracted from the model without ongoing engagement with communities. 

Another important concern raised by researchers was the potential damage that using LLMs could inflict on the trust between qualitative researchers and the communities they engage with. Historically, many vulnerable communities have developed deep distrust toward researchers due to exploitative practices in which academics extracted data or introduced interventions without offering ongoing support or maintenance. This legacy of distrust could be further exacerbated if researchers begin to replace participant perspectives with LLM-generated responses. Yue, who works closely with d/Deaf participants, expressed concern that such practices would erode trust in research, especially within communities that are already wary of being misrepresented. Reducing these communities' voices to algorithmic outputs undermines the value of their lived experiences, potentially eroding the trust that researchers worked hard to build.

The concerns expressed by researchers reflect a broader anxiety about the potential implications of relying on LLMs in qualitative research. For Nikita, the use of LLMs evoked a \textit{``sense of the dystopian,''} where, for example, trans communities are not heard by legislators or other decision-makers, as their perspectives are distorted by technology rather than being directly represented. Cameron summarized this concern by emphasizing that such tools miss the \textit{``epistemological underpinnings of qualitative methods and why we do them''}. She argued that if the goal of qualitative research is to obtain contextualized data that is grounded in people's lived experiences, then LLMs fundamentally fail to meet this standard. LLMs may generate text that \textit{appears} fluent and contextually relevant, but this output lacks the depth, nuance, and authenticity that come from direct engagement with people's lived realities. 

\subsection{Imagining Use Cases for LLMs in Qualitative Research: Seeing Possibilities and Even More Caveats}
\label{use}
Most researchers voiced their discomfort with using LLMs to generate synthetic research data, with several even suggesting that they would distrust findings from studies that utilize LLMs as research participants. As a thought exercise, we explored whether there might be specific domains, contexts, or use cases where LLMs could be more effective. Below, we outline the use cases that researchers identified as \textit{more} appropriate than replacing direct engagement with communities. It is important to note that our participants did not reach a consensus on any of these use cases. For each potential application, participants also highlighted ways in which LLM use could be harmful or ineffective.

Interviewees suggested that using LLMs to simulate participants could serve as a valuable pedagogical tool where the stakes are lower than in a research study. This approach could help novice researchers learn how to focus on specific aspects of responses and ask effective follow-up questions. Some interviewees also highlighted flaws in this proposition, noting that prompting an LLM often differs significantly from interacting with human participants. For example, Daria had to prompt the model thrice before it provided enough contextual detail to make a useful response. Researchers emphasized that a critical interviewing skill is managing emotions—their own and the participant's—which is difficult to replicate when learning qualitative methods with LLMs. Jasmine also pointed out the risk that junior researchers learning qualitative methods with LLMs might develop poor interviewing habits (\eg not building rapport, not probing for depth, interrupting participants, or ignoring body language) that do not translate well to real research contexts and could perpetuate the notion of \textit{``extracting information''} from research subjects.

For most researchers, LLMs could, at their best, help test their interview protocol, especially when it is challenging to recruit participants for pilot studies. In those situations, LLMs could serve as stand-ins, allowing researchers to gauge the types of responses a question might elicit or determine what kinds of questions to ask. Here, Mario cautioned that relying on LLMs for pilot interviews or refining interview guides could shift the focus of the project in unexpected ways. He pointed out how LLMs might lead researchers to become overly fixated on specific directions that might otherwise not occur if researchers were interviewing real participants. 

Several researchers observed that the decision to use LLMs depends on the research topic and the community being simulated. In sensitive research contexts, such as those involving experiences of oppression or discrimination, participants noted that engaging with LLMs could carry some potential benefits and the significant risks described above. Sensitive research topics risk placing an epistemic burden on participants and re-traumatizing them by asking them to revisit difficult experiences. While some researchers suggested that LLMs could be useful in these contexts, they also expressed concerns that using LLMs for sensitive topics might exacerbate the erasure of real human experiences. Nadia, for example, was skeptical about the ability of LLMs to accurately simulate complex human experiences, such as navigating gender identity or sexuality in the workplace or the experiences of refugees and migrants. On the other hand, Nadia expressed how there are certain communities, such as hate groups, that they don't feel safe engaging with. She argued that those interviews could be simulated with an LLM to help develop defensive strategies against hate groups online. Nikita argued that for sensitive topics, researchers should consider collaborating with community members to gain relevant expertise and learn how to have difficult conversations, rather than \textit{``turning to an LLM.''} % they also stressed the importance of careful training and thoughtful engagement when working with vulnerable populations.

\section{Discussion}

\begin{aquote}{Neda Atanasoski and Kalindi Vora \cite{atanasoski2019surrogate}}
    The claim that technologies can act as surrogates recapitulates histories of disappearance, erasure, and elimination necessary to maintain the liberal subject as the agent of historical progress. 
\end{aquote}

We argue that the use of LLMs as research subjects enacts what Atanasoski and Vora describe as the `surrogate effect' \cite{atanasoski2019surrogate}. In \textit{Surrogate Humanity}, they reveal how technologies like social robots and artificial intelligence are not neutral tools but are instead shaped by capitalist logics of differential exploitation and dispossession. These technologies are designed to act as surrogates, ostensibly to emancipate humans from ``historically degraded tasks'' that are considered less-than-human. When LLMs are used as stand-ins for human participants in research, they \textit{similarly} function as surrogates--replacing the voices and lived experiences of actual communities with algorithmic approximations.

It is crucial to acknowledge the invisible gendered, classed, and racialized labor underpinning these systems.
LLMs are not self-sustaining entities that emerge in a vacuum; their creation involves vast networks of a global workforce that contribute to the training and maintenance of these systems \cite{Miceli2022The}. These workers, who are often underpaid and overworked \cite{Sambasivan2021EveryoneWT} engage in the labor of labeling data, evaluating and moderating content, and correcting model errors, all of which are essential for the functionality of LLMs. Tech companies and platforms aim to disguise this human labor as machine labor to create a veneer of machine intelligence \cite{atanasoski2019surrogate}. As Atanasoski and Vora articulate, "labor becomes something that is intentionally obfuscated to create the effect of machine autonomy" \cite{atanasoski2019surrogate}. This intentional obfuscation creates a \textit{conjuration of algorithms} \cite{Nagy2024ConjuringAU} that enables the fantasy of magical techno-objects and masks the exploitation and uneven distribution of labor benefits in the global tech industry. 

Science and Technology Studies (STS) scholars have long interrogated the ways in which technologies are not merely tools--for research or otherwise-- but are imbued with social, political, and economic values. Sheila Jasanoff’s concept of \emph{co-production} highlights how scientific knowledge and social order are produced together, each shaping and stabilizing the other \cite{Jasanoff2004States}. LLMs' role as surrogates in research is also a co-production of knowledge and power. These models are encoded with the biases, assumptions, and power dynamics of model producers and the data and contexts from which they are derived. As such, their use in research does not merely replace human participants; it reshapes the nature of the knowledge produced, often in ways that reinforce existing hierarchies and exclusions. 

We advocate for the need to critically examine how LLMs, products of specific socio-technical assemblages, contribute to a reconfiguration of what counts as valid knowledge and who gets to produce it. Technological artifacts have politics, too, as they embody specific forms of power and authority \cite{Winner2017Do}. The use of LLMs as research subjects exemplifies this notion by centralizing the role of technology in knowledge production while marginalizing the participation of human subjects, particularly those from vulnerable or historically underrepresented communities. Below, we draw on scholarship from qualitative epistemologies and research ethics \& care to argue for a critical reconsideration of the use of LLMs as surrogate participants, advocating for caution and, in most cases, avoiding such practices.

\subsection{LLMs are incongruent with qualitative epistemologies}

Human subjects research is an active process of constructing meaning that Ithiel de Sola Pool describes as an \textit{``interpersonal drama with a developing plot''} \cite{Silverman2004QualitativeR}. Qualitative research epistemologies, especially within HCI, emphasize the importance of researcher positionality: the idea that knowledge is co-constructed by the researcher and the participant and is deeply influenced by the specific context in which it is generated. The interview is not a neutral process of extracting information that exists \textit{out there} \cite{10.1145/1978942.1979042}; instead, research interviews are intentional, and the interviewer is implicated in the process of assembling knowledge. Research participants, too, are not \textit{`vessels of answers'} waiting to be tapped in to elicit truths about facts, opinions, or their lived experiences \cite{law2004after}. Such an interpretivist orientation inherently values the nuances, complexities, and particularities of lived experience, often best captured through methods that allow for deep engagement, reflection, and interpretation.

Indeed, there are many sites of knowledge production, particularly in a community-based project, that exist outside the confines of a survey questionnaire or one-hour interview \cite{van2011tales}. This multisitedness, articulated by Law \cite{law2004after}, involves a multiplicity of locations where \textit{`analysis and knowledge are made material.'} LeDantec and Fox \cite{ledantec2015strangers} describe this as work that comes before the work: building relationships and demonstrating commitments that shape the outcomes of the projects. We argue that interviewing with LLMs is fundamentally misaligned with the notion that \textit{``methods [no longer] discover and depict realities. they participate in the enactments of those realities.''} \cite{law2004after}

Finally, qualitative methods are designed to produce knowledge that is not only rich and detailed, in the form of a \textit{thick description} \cite{geertz2008thick}, but also situated within specific cultural, social, and material contexts. Haraway's idea of situated knowledges invites a critical perspective that helps to problematize \textit{`the god trick of seeing everything from nowhere'} \cite{haraway2013situated}. Haraway urges us to resist myths of infinite vision, seemingly offered by the visualization tricks and powers of modern technology. In thinking with Haraway, we contend that the prevailing logics of LLMs as proxies for human behavior offer similar \textit{unlocatable} knowledge claims. 

LLMs, by their very design, operate on principles that are at odds with feminist epistemological foundations. They are built to extract patterns that can be applied across contexts with the aim of achieving universality and objectivity \cite{Barocas2018FairnessAM}. While this capacity can seemingly lead to efficiencies and broad applicability, it also means that the knowledge produced by LLMs is often decontextualized and disembodied. The models are trained on data that, despite its volume, may lack the depth and specificity that qualitative researchers seek. Moreover, LLMs do not account for the positionality of the researcher or the participants; instead, they flatten differences and obscure the peculiarities that are central to qualitative inquiry. This turn to LLMs reflects a world in which the complexity and messiness of human experience are reduced to manageable data points that can be easily processed by algorithms. In doing so, the use of LLMs in research aligns with a technocratic vision of science that prioritizes efficiency and scalability over contextual engagement with communities. 

\subsection{LLM use undermines consent and autonomy of data subjects}

HCI research has long been dedicated to supporting the autonomy and agency of research participants. We recognize the importance of ensuring that individuals have control over their participation and the information they share. In the broader context of coming to a shared understanding of research ethics, HCI and CSCW scholars have expressed concerns with the use of `public' social media data for research purposes \cite{klassen2022isn, zimmer2020but}. This includes practices, such as the mining of profile pictures, to identify social categories such as sexual orientation \cite{levin2017new} or facial recognition from students' photos \cite{fussell2019you}. Social media users are often unaware that their data is being used by researchers, and in fact, many participants in Fiesler and Proferes's study asserted that researchers should not be able to use tweets without explicit consent \cite{fiesler2018participant}. Online content creators, including social media users, may not realize the public nature of their data and the downstream applications the data may be supporting \cite{fiesler2018participant}. 
% what are risks of public data???? 

We believe similar risks apply when researchers use LLMs to create synthetic interview data. Most current LLMs are trained on public content extracted from the Internet \cite{byrd2023truth}. Earlier this year, Google and OpenAI signed contracts with Reddit that give these companies real-time access to Reddit posts \cite{googlereddit, openaireddit}. OpenAI maintains that `training AI models using publicly available internet materials is fair use, as supported by long-standing and widely accepted precedents' \cite{openAIpublicdata}. Many other big tech companies have made arguments to similar effect when faced with lawsuits. While training on publicly available and/or copyrighted text may not be unlawful (yet), generating research data using LLMs--models created through the expropriation of the work of others--presents the risk of undermining the autonomy of data subjects powering these applications. 

A productive framework for understanding the relationship between autonomy and consent in research comes from Beauchamp's work \cite{beauchamp2010autonomy}, which challenges the traditional notion of `informed consent.' Rather than focusing narrowly on the disclosure of information to research subjects, Beauchamp advocates for a more nuanced understanding of `voluntary consent' that includes self-authorization. We call for extending this consideration beyond our direct interactions with research subjects to the individuals whose data is used to produce these models, particularly in terms of how their autonomy is respected--or potentially subverted. For example, researchers using LLMs to simulate participants might expose sensitive information from online communities, such as discussions on Reddit about particular worker resistance tactics that would have remained largely unknown otherwise. These communities, whose members might not have intended to share their experiences with researchers, could find their information under a microscope without their explicit consent. 

\subsection{What does it mean to simulate the user in HCI?}

One might contend that HCI's primary goal is to understand how people interact with technology and design systems that meet their needs \cite{olson2014ways}. Any attempts to use simulations, such as user personas or LLMs, to represent communities' interests must then be critically examined. Design Justice, as articulated by scholars like Sasha Costanza-Chock \cite{costanza2020design}, emphasizes the full inclusion of people with direct lived experience throughout the design process. Chock raises the fundamental question: ``For whom do we design technology?'' If our systems are designed to work only for the `unmarked' group (those who are White, male, heterosexual, able-bodied, and literate), since simulations tend to reflect dominant perspectives, we risk reinforcing the spiral of exclusion. 
A clear example of this is disability simulation, where designers attempt to `step into the shoes' of disabled individuals. Disability scholars have long critiqued this approach for producing unrealistic and reductive understandings of life with a disability \cite{silverman2015stumbling, bennett2019promise}. Instead, HCI should be guided by the principle of `nothing about us without us,' central to many social justice movements \cite{spiel2020nothing}. This principle goes beyond token participation: it requires deep engagement with, accountability towards, and ownership by those impacted by the systems we introduce.

Our participants were acutely aware of the misalignment between these principles and the use of large language models as surrogate participants. They suggested, instead, potential scenarios where using LLMs within the research process might be more appropriate than using them to substitute participation from research communities. This includes using LLMs to trial materials, as a pedagogical device, and in sensitive contexts or unsafe environments. Researchers also went on to articulate concerns with each of these use cases, such as erasure and misrepresentation or misleading qualitative research training. %It is crucial to recognize that such use cases do not bypass the fundamental limitations of using LLMs in qualitative research. 
Below, we outline three critical considerations (transparency, model selection, and validation) for researchers who find appeal in the use of LLMs as simulated qualitative research participants. These considerations are not meant to serve as guidance to justify the use of LLMs. On the contrary, we urge careful reflection on how the use of LLMs involves complex methodological choices that significantly impact the integrity of qualitative research. 

Firstly, thoughtful model selection is not merely a performance-driven decision but one that directly impacts the validity of the synthetic research data. For example, a model trained predominantly on Western-centric datasets may not accurately simulate experiences from non-Western contexts \cite{khandelwal2024indian}. 
Different models undergo different alignment processes depending on the goals and values of the developers. Unless there is strong evidence to believe that data \textit{from} a minority group (not about them) is included in the training data, researchers should exercise caution when drawing inferences from simulated interview data. Secondly, researchers must be fully transparent about any use of LLMs in their studies, in line with the ACM policy \cite{ACMPolicy}, including any pilot studies with simulated participants. 
However, simply disclosing information does not automatically invite critique or enable a better understanding of the research design; instead, visibility often shifts the burden onto individuals. Transparency can create an illusion of accountability without fostering meaningful engagement or critical reflection. Transparency, then, risks becoming \emph{performative}, especially if disclosure can only be selective. 

Finally, it would be crucial to engage in rigorous validation processes, such as member checking, where findings or interpretations are returned to community members for review and feedback \cite{birt2016member}. When LLMs are involved, member checking introduces additional layers of complexity. Engaging with simulated data that reflects one’s community or personal experiences demands significant time, energy, and emotional investment. Reading simulated experiences, especially those that may inaccurately represent or trivialize lived experiences, could be distressing for community members. Encouraging community members to critically engage with LLM-generated data and voice disagreements if the synthetic data does not align with their experiences could also prove challenging. Researchers must be prepared to re-evaluate the LLM's outputs, revise initial interpretations, or even reconsider the appropriateness of using LLMs if the feedback indicates misalignment with the community’s perspectives. It is important to recognize that none of these considerations address the deeper epistemological questions with the use of LLMs.
\section{Conclusion}

This paper focuses on examining the use of large language models as proxies for research participants. This is no longer a speculative future; research papers and various products are already exploring LLMs' ability to replace human participants. Through interviews with qualitative researchers, we highlight several significant limitations of using LLMs as stand-ins for qualitative research participants, including the lack of palpability in LLM responses, the ambiguity surrounding the model's epistemic position, the amplification of researcher positionality, how the use of LLMs forecloses participants' consent and agency, the erasure of community perspectives, and the risk of undermining qualitative methods of knowing. We conclude by drawing on STS and research ethics literature to discuss how these practices bring to light critical questions about consent, agency, and the legitimacy of using LLMs in qualitative research. The ethical and epistemological tensions demand a cautious and, in many cases, critical stance toward this form of substitution in research that seeks to understand lived experiences.

%\input{Sections/Acknowledgements}

% \begin{acks}
% \end{acks}

%%\input{Sections/Introduction}
\bibliographystyle{ACM-Reference-Format}
\bibliography{references}

\end{document}